\documentclass[acmsmall]{acmart}

\usepackage{graphicx}
\usepackage{textcomp}
\usepackage{xcolor}
\usepackage{ifthen}
\usepackage{booktabs}
\usepackage{scalefnt}
\usepackage{multirow}
\usepackage{color, colortbl}
\usepackage[flushleft]{threeparttable}
\usepackage{caption}
\usepackage{subcaption}
\usepackage{pifont}

\newboolean{showcomments}
\setboolean{showcomments}{true} 
\ifthenelse{\boolean{showcomments}}
 {
  
 }
 {\newcommand{\nbb}[2]{}
  
 }
 
\newcommand{\MyBox}[1]{\vspace{4mm}\noindent\framebox[\columnwidth][c]{\parbox[b]{0.95\columnwidth}{ #1 }}\vspace{4mm}}

\AtBeginDocument{%
  \providecommand\BibTeX{{%
    \normalfont B\kern-0.5em{\scshape i\kern-0.25em b}\kern-0.8em\TeX}}}

\setcopyright{acmcopyright}
\copyrightyear{2021}
\acmYear{2021}
\acmDOI{10.1145/1122445.1122456}

\acmJournal{JACM}
\acmVolume{37}
\acmNumber{4}
\acmArticle{111}
\acmMonth{8}

\newcommand*{\img}[1]{%
    \raisebox{-.3\baselineskip}{%
        \includegraphics[
        height=\baselineskip,
        width=\baselineskip,
        keepaspectratio,
        ]{#1}%
    }%
}

\begin{document}


\title{Don't Disturb Me: Challenges of Interacting with Software Bots on Open Source Software Projects}

\author{Mairieli Wessel}
\affiliation{\institution{University of Sao Paulo, Brazil}}
\email{mairieli@ime.usp.br}

\author{Igor Wiese}
\affiliation{\institution{Universidade Tecnologica Federal do Parana, Brazil}}
\email{igor@utfpr.edu.br}

\author{Igor Steinmacher}
\affiliation{\institution{Universidade Tecnologica Federal do Parana, Brazil}}
\email{igorfs@utfpr.edu.br}

\author{Marco A. Gerosa}
\affiliation{\institution{Northern Arizona University, USA}}
\email{marco.gerosa@nau.edu}
\renewcommand{\shortauthors}{Mairieli Wessel et al.}

\begin{abstract} 
Software bots are used to streamline tasks in Open Source Software (OSS) projects' pull requests, saving development cost, time, and effort. However, their presence can be disruptive to the community. We identified several challenges caused by bots in pull request interactions by interviewing 21 practitioners, including project maintainers, contributors, and bot developers. In particular, our findings indicate noise as a recurrent and central problem. Noise affects both human communication and development workflow by overwhelming and distracting developers. Our main contribution is a theory of how human developers perceive annoying bot behaviors as noise on social coding platforms. This contribution may help practitioners understand the effects of adopting a bot, and researchers and tool designers may leverage our results to better support human-bot interaction on social coding platforms.
\end{abstract}



\maketitle

\section{Introduction}

Open Source Software (OSS) development is inherently collaborative, frequently involving geographically dispersed contributors. OSS projects often are hosted in social coding platforms, such as GitHub and GitLab, which provide features that aid collaboration and sharing, such as pull requests~\cite{Tsay:2014:LTE}. Pull requests facilitate interaction among developers to review and integrate code contributions. To alleviate their workload~\cite{Gousios.Bachelli_2016}, project maintainers often rely on software bots to check whether the code builds, the tests pass, and the contribution conforms to a defined style guide~\cite{bogdan2015,Gousios:2015:ICSE,choicesmatter}. More complex tasks include repairing bugs~\cite{urli2018design,Monperrus2019}, refactoring source code~\cite{Wyrich2019}, recommending tools~\cite{Brown2019}, updating dependencies~\cite{mirhosseini2017can}, and fixing static analysis violations~\cite{botc3pr}.

The introduction of bots aims to save cost, effort, and time~\cite{Storey2016}, allowing maintainers to focus on development and review tasks. However, new technology often brings consequences that counter designers' and adopters' expectations~\cite{healy2012unanticipated}. Developers who \emph{a priori} expect technological developments to lead to performance improvements can be caught off-guard by \emph{a posteriori} unanticipated operational complexities and collateral effects~\cite{woods2001unexpected}. For example, the literature has shown that although the number of human comments decreases after the introduction of bots~\cite{wessel2020effects}, many developers do not perceive this decrease~\cite{Wessel2020whatexpect}. These collateral effects and the misalignment between the preferences and needs of project maintainers and bot developers can cause expectation breakdowns, as illustrated by a developer: ``\textit{Whoever wrote <bot-name> fundamentally does not understand software development.}''\footnote{\url{https://twitter.com/mojavelinux/status/1125077242822836228}} Moreover, as bots have become new voices in developers' conversation~\cite{Monperrus2019}, they may overburden developers who already suffer from information overload when communicating online~\cite{nematzadeh2016information}. On an abandoned pull request, a maintainer complained about the frequency of action of a bot: ``\textit{@<bot-name> seems pretty active here [...].}''\footnote{\url{https://github.com/facebook/react/pull/12457\#issuecomment-413429168}} Changes the technology provokes in human behavior may cause additional complexities~\cite{mulder2013impact}. Therefore, it is important to assess and discuss the effects of a new technology on group dynamics; yet, this is often neglected when it comes to software bots~\cite{Storey2016,Paikari.vanDerHoek:2018}.

Considering developers' perspectives on the overall effects of introducing bots, designers can revisit their bots to better support the interactions in the development workflow and account for collateral effects. So far, the literature presents scarce evidence, and only as secondary results, of the challenges incurred when adopting bots. Investigating the usage of the \textit{Greenkeeper} bot, \citet{mirhosseini2017can}, for example, report that maintainers are overwhelmed by bot pull request notifications interrupting their workflow.
According to \citet{Brown2019}, the human-bot interaction on pull requests can be inconvenient, leading developers to abandon their contributions. This problem may be especially relevant for newcomers, who require special support during the onboarding process due to the barriers they face~\cite{Steinmacher:2015:SBF:2675133.2675215,Steinmacher.Conte.ea_2016}. Newcomers can perceive bots’ complex answers as discouraging, since bots often provide a long list of critical contribution feedback (e.g., style guidelines, failed tests), rather than supportive assistance.

We extend previous work by delving into the challenges incurred by bots on social coding platforms. To do so, we investigate the challenges bots bring to the pull request workflow from the perspective of practitioners.
Specifically, our work investigates the following research question:

\begin{itemize}
    \item[\textbf{RQ.}] \textit{What interaction challenges do bots introduce when supporting pull requests?}
\end{itemize} 

To answer our research question, we qualitatively analyzed data collected from semi-structured interviews with 21 practitioners, including OSS project maintainers, contributors, and bot developers who have experience interacting with bots on pull requests. After analyzing the interviews, we validated our findings through member-checking.

While participants commend bots for streamlining the pull request process, they complain about several challenges they introduce. Some of the challenges include annoying bot behaviors such as verbosity, too many actions, and unrequested or undesirable tasks on pull requests, which are often perceived as noise. Since noise emerged as a central theme in our analysis, we further theorize about it, grounded in the data we collected. 

Our work contributes to the state-of-the-art by (i) identifying a set of challenges incurred by the use of software bots on the pull requests' workflow and (ii) proposing a theory about how noise introduced by bots disrupts developers' communication and workflow. 
By gathering a comprehensive set of challenges incurred by bots, our findings complement the previous literature, which presents scarce and diffuse challenges, reported as secondary results. Our contributions support practitioners in understanding, or even anticipating, the impacts that adopting a bot may have on their projects. Researchers and tool designers may also leverage our results to enhance bots' communication design, thereby better supporting human-bot interaction on social coding platforms.

\section{Background}
\label{sec:literature-problems}

According to \citet{Storey2016}, a software bot is \textit{``a conduit or an interface between users and services, typically through a conversational user interface''}. In the following, we provide more details about the existing literature related to the challenges of using software bots, especially on social coding platforms. 

\subsection{Challenges of bots in online communities}

Software bots have been extensively studied in the literature of different domains, including social media~\cite{savage2016botivist,abokhodair2015dissecting,xu2014sobot,ferrara2016rise}, online learning~\cite{ghose2013toward,latham2010oscar,nakamura2012investigation}, and Wikipedia~\cite{geiger2017operationalizing,cosley2007suggestbot}. Despite the widespread adoption of bots in different domains, the interaction between computers and humans still presents challenges~\cite{dale2016return,vinciarelli2015open,zue2000conversational}. 
For example, \citet{zheng2018steps} describe how although editors appreciate Wikipedia bots for streamlining knowledge production, they complain that the bots create additional challenges. To circumvent some of these challenges, Wikipedia established rigid governance roles~\cite{muller2013work}. Bots need to contain the string ``bot'' in their username, have a discussion page that clearly describes what they do, and can be turned off by any member of the community at any time.

In recent years, software bots have also been proposed to support collaborative software engineering, encompassing both technical and social aspects of software development activities~\cite{lin2016developers}. According to \citet{lebeuf2018software}, bots provide an interface with additional value on top of the software service's basic capabilities. Interviewing industry practitioners, \citet{erlenhov2020empirical} found that bots cause interruption and noise, trust, and usability issues. 

\subsection{Challenges of bots on social coding platforms}

In the scope of social coding platforms, \citet{Wessel2018} conducted a study to characterize the bots that support pull requests on GitHub. Their results indicate that bot adoption is widespread in OSS projects and are used to perform a variety of tasks on pull requests. The authors also report some challenges of using bots on pull requests. Several contributors complained about the way the bots interact, saying that the bots provide non-comprehensive or poor feedback. In contrast, others mentioned that bots introduce communication noise and that there is a lack of information on how to interact with the bot. 

Certain bots have been studied in detail, revealing challenges and limitations of their interventions in pull requests. For example, while analyzing the \textit{tool-recommender-bot}, \citet{Brown2019} report that bots still need to overcome problems such as notification workload. \citet{mirhosseini2017can} analyzed the \textit{greenkeeper} bot and found that maintainers were often overwhelmed by notifications and only a third of the bots' pull requests were merged into the codebase. \citet{Peng2019} conducted a case study on how developers perceive and work with \textit{mention bot}. The results show that this bot has saved developers' efforts; however, it may not meet the diverse needs of all users. For example, while project owners require simplicity and stability, contributors require transparency, and reviewers require selectivity. Despite its potential benefits, results also show that developers can be bothered by frequent review notifications when dealing with a heavy workload.

Although several bots have been proposed, relatively little has been done to evaluate the state of practice. Furthermore, although some studies focus on designing and evaluating bot interactions, they do not draw attention to potential problems introduced by these bots at large. According to \citet{Brown2019}, bots still need to enhance their interaction with humans. Responding to this gap, we complement the findings from previous works by delving deeper into the challenges that bots bring to interactions on social coding platforms. This study takes a closer look at how practitioners interact with bots and what challenges they face. Also complementing the previous literature, we discuss how noise is characterized in terms of its impacts and how developers have attempted to handle it.

\section{Research Design}

\begin{figure*}[!bthp]
\scriptsize
\centering
\includegraphics[width=0.7\textwidth]{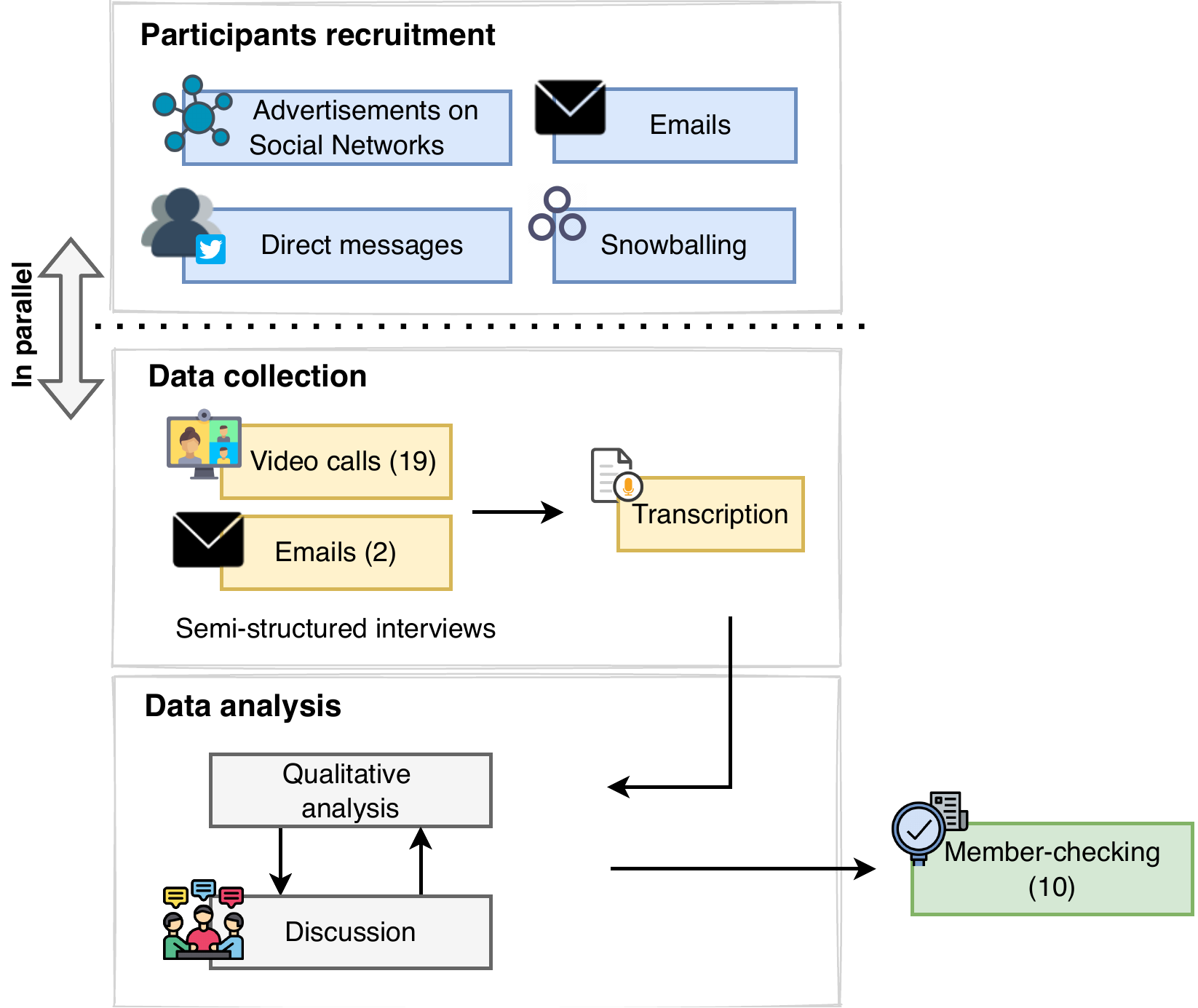}
\caption{Research Design Overview}
\label{fig:overview}
\end{figure*}

The main goal of this study is to identify challenges caused by bots on pull request interactions. To achieve this goal, we conducted a qualitative study of responses collected from semi-structured interviews. Figure~\ref{fig:overview} shows an overview of the research design employed in this study.

\subsection{Participants recruitment}

We recruited participants from three different groups: (i) project maintainers, (ii) project contributors, and (iii) bot developers. Participants were expected to have experience contributing to or maintaining projects that use bots to support pull request activities. We adopted four main strategies to invite participants: (i) advertising on Twitter, (ii) direct messages, (iii) emails, and (iv) snowballing. Besides the broad advertisement posted on Twitter, we also manually searched for users that had posted about GitHub bots or commented on posts related to GitHub bots. During this process, we sent direct messages to 51 developers. In addition, we used the GitHub API to collect a set of OSS projects that use more than one bot. After collecting a set of 225 GitHub repositories using three or more bots, we sent 150 emails to maintainers and contributors whose contact information was publicly available. In addition, we asked participants to refer us to other qualified participants. We continued inviting participants as the data unveiled new relevant information. The participants received a 25-dollar gift card as a token of appreciation for their time. 

\begin{table}[!htbp]
\centering
\caption{Demographics of interviewees}
\label{tab:participants}
\begin{threeparttable}
\scriptsize
\begin{tabular}{cccccc}
\toprule
\textbf{Participant} & \textbf{SD Experience} & \multicolumn{3}{c}{\textbf{Experienced with bots as}}  & \textbf{Location} \\
\textbf{ID} & \textbf{(years)} &  \textbf{Maintainer} & \textbf{Contributor} & \textbf{Bot developer} & \\
\midrule
P1 & 9 & \ding{51} & \ding{51} & & Europe \\ \hline
P2 & 2 & \ding{51} & & & South America  \\ \hline
P3 & 20 & \ding{51} & \ding{51} & \ding{51} & North America \\ \hline
P4 & 10 & \ding{51} & & \ding{51} & North America \\ \hline
P5 & 12 & \ding{51} & \ding{51} & \ding{51} & North America \\ \hline
P6 & 4 & \ding{51} & \ding{51} & & North America \\ \hline
P7 & 10 & \ding{51} & & & North America \\ \hline
P8 & 10 & \ding{51}* & & & North America \\ \hline
P9 & 14 & \ding{51} & & \ding{51} & Europe \\ \hline
P10 & 12 & \ding{51} & \ding{51} & & South America \\ \hline
P11 & 5 & & \ding{51} & & Europe \\ \hline
P12 & 20 & \ding{51} & & \ding{51} & North America \\ \hline
P13 & 25 & \ding{51} & &  & North America \\ \hline
P14 & 25 & \ding{51} & & & Europe \\ \hline
P15 & 13 & & \ding{51} & & North America \\ \hline
P16 & 20 & \ding{51} & \ding{51} &  & Europe \\ \hline
P17 & 8 & \ding{51} & & \ding{51} & North America \\ \hline
P18 & 5 & & \ding{51} & & Europe \\ \hline
P19 & 5 & \ding{51} & \ding{51} & & North America\\ \hline
P20 & 4 & \ding{51} & & & Europe \\ \hline
P21 & 11 & \ding{51} & & & Europe \\
\bottomrule 
\end{tabular}
\begin{tablenotes}
 \item * Also described himself as a \textit{bot evangelist}
\end{tablenotes}
\end{threeparttable}
\end{table}

As a result of our recruitment, we interviewed 21 participants---identified here as P1 -- P21. Table~\ref{tab:participants} presents the demographics of our interviewees. Their experience with software development ranges from 3 to 25 years ($\simeq$ 12 years on average). Participants are geographically distributed across North America ($\simeq$53\%), Europe ($\simeq$38\%), and South America ($\simeq$10\%). 
Three interviewees are project contributors who have interacted with bots when submitting pull requests to open-source projects. All the other interviewees (18) maintain projects that use bots to support pull request activities. Besides their experience as project maintainers, seven of them also have experience in contributing to other projects that use bots. Six maintainers have experience building bots. One of the maintainers also described himself as a ``bot evangelist.''

Additionally, participants have experience with diverse types of bots, including project-specific bots, dependency management bots (e.g., Dependabot, Greenkeeper), code review bots (e.g., Codecov, Coveralls, DeepCode), triage bots (e.g., Stale bot), and welcoming bots (e.g., First Timers bot). Their experience ranges from interacting with 1 to 6 bots ($\simeq$ 2 bots on average), encompassing a total of 24 different bots. Further, bot developers develop between 1 and 3 bots ($\simeq$ 1 on average). For confidentiality reasons, we do not report either the bots used by each participant or their projects.

\subsection{Semi-structured interviews}

To identify the challenges, we conducted semi-structured interviews, which comprised open- and closed-ended questions that enabled interviewers to explore interesting topics that emerged during the interview~\cite{hove2005experiences}. By participants' requests, 2 interviews (P1 and P20) were conducted via email. The other 19 interviews were conducted via video calls. We started the interviews with a short explanation of the research objectives and guidelines, followed by demographic questions. The rest of the interview script focused on three main topics: (i) experience with GitHub bots, (ii) main challenges introduced by the bots, and (iii) the envisioned solutions to those challenges. The detailed interview script is publicly available\footnotemark. Each interview was conducted remotely by the first author of this paper and lasted, on average, 46 minutes. 

\subsection{Data analysis}

We qualitatively analyzed the interview transcripts, performing open and axial coding procedures~\cite{strauss1998basics,stol2016grounded} throughout multiple rounds of analysis. We started by applying open coding, whereby we identified challenges brought by the interaction, adoption, and development of bots. To do so, the first author of this paper conducted a preliminary analysis, identifying the main codes. Then, we discussed the coding in weekly hands-on meetings, aiming to increase the reliability of the results and mitigate bias~\cite{Strauss.Corbin_1998, patton2014qualitative}. In these meetings, all the researchers revisited codes, definitions, and their relationships until reaching an agreement. Afterwards, the first author further analyzed and revised the interviews to identify relationships between concepts that emerged from the open coding analysis (axial coding). Then, the entire group of researchers discussed the concepts and their relationships during the next weekly meeting. During this process, we employed a constant comparison method~\cite{glaser2017discovery}, wherein we continuously compared the results from one interview with those obtained from the previous ones. The entire analysis lasted eight weeks and each weekly meeting lasted from 1 to 2 hours.

For confidentiality reasons, we do not share the interview transcripts. However, we made our complete code book publicly available\footnotemark[\value{footnote}]. 
The code book includes the code names, descriptions, and examples of quotes for all categories.

\footnotetext{\url{https://zenodo.org/record/4088774}}

\subsection{Member-checking}

As a measure of trustworthiness, we member-check our final interpretation of the theory about noise introduced by bots with the participants. The process of member-checking is an opportunity for participants check particular aspects of the data they provided~\cite{merriam1998qualitative}. According to Charmaz~\cite{charmaz2006constructing}, member-checking entails ``\textit{taking ideas back to research participants for their confirmation.}'' Such checks might occur through returning emerging research findings or a research report to individual participants for verification of their accuracy. 

We contacted our 21 participants via email. In the email, we included the theory, followed by a short description of the concepts and their relationships. Participants could provide feedback by email or through an online meeting. Ten participants provided their feedback: P2, P3, P4, P7, P13, P16, P18, and P20 provided a detailed feedback by email, whereas P10 and P12 scheduled an online meeting, each lasting about 20 minutes. 
 
The participants who gave feedback agreed with the accuracy of the theory about noise introduced by bots. P4, an experienced bot developer, described our research in a positive light, saying it ``\textit{captures the problem of writing an effective bot.}'' The participants suggested a few adjustments. For instance, P12 recommended including another countermeasure to avoid noise (``\textit{re-designing the bot''}). We addressed the feedback by including this countermeasure to our theory. Additional comments from member-checking can be found in our code book, tagged as ``\textit{from member-checking}''.

\section{Findings}

In this section, we present the challenges reported by the participants, as well as a theory focused on explaining the reasons and effects of the noise caused by bots on pull requests.

\subsection{Challenges incurred by bots}
\label{sec:general-problems}

\begin{figure*}[!bthp]
\scriptsize
\centering
\includegraphics[width=\textwidth]{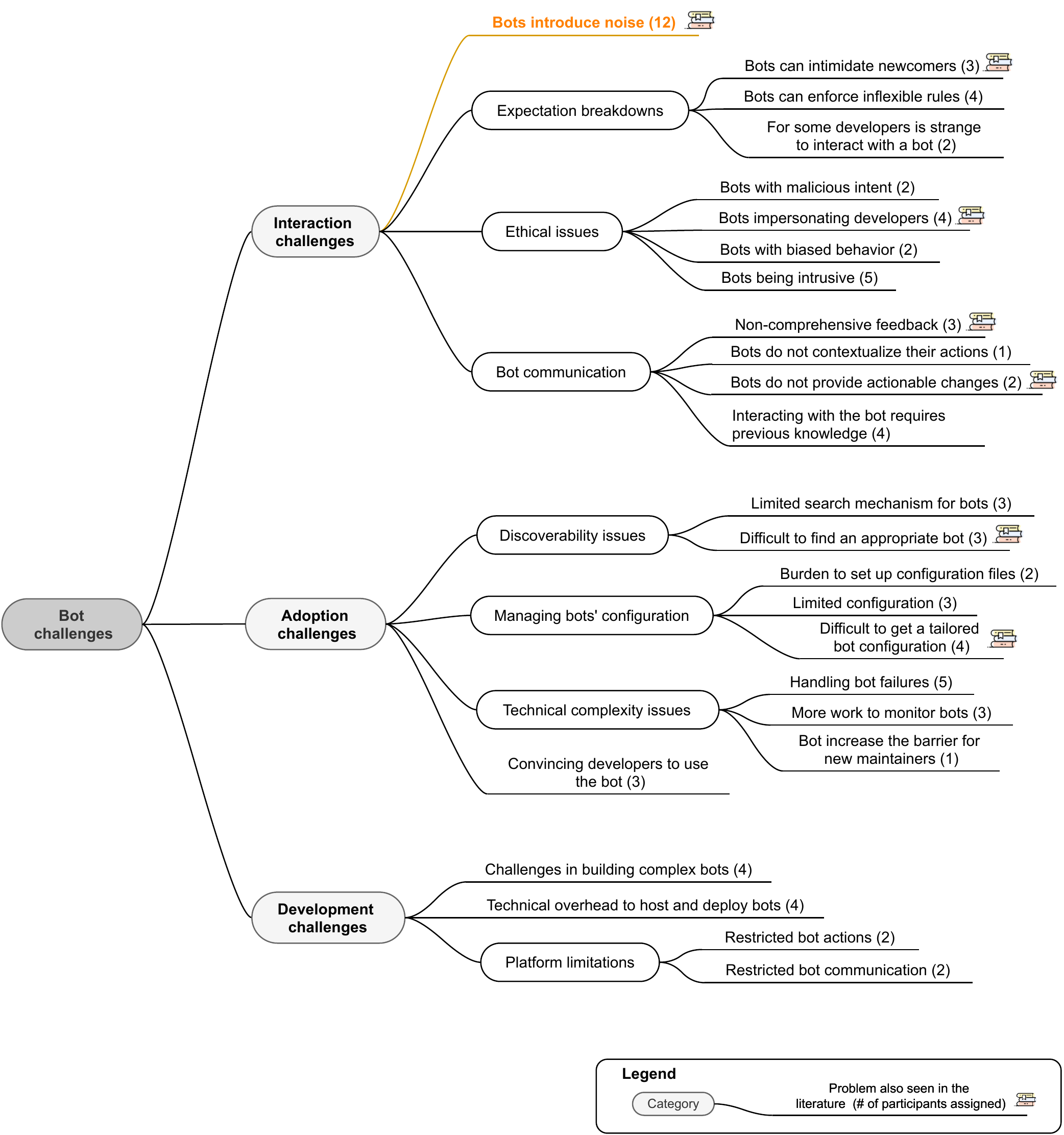}
\caption{Hierarchical Categorization of Bot Challenges}
\label{fig:research}
\end{figure*}

The interviewees reported social and technical challenges related to the development, adoption, and interaction of bots on pull requests. Figure~\ref{fig:research} shows a hierarchical categorization that summarizes these challenges.
We added a graphical mark (\img{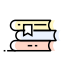}) in the hierarchical categorization to identify challenges that have been also identified by previous work, described in Section~\ref{sec:literature-problems}. In summary, we found 25 challenges, organized into three categories (development challenges, adoption challenges, and interaction challenges) and several sub-categories. In the following, we present these three main categories of challenges, focusing on the 12 challenges related to the human-bot interaction on pull requests, since they strongly align with the challenges posed by the theory about noise introduced by bots. We describe the categories in \textbf{bold}, and provide the number of participants we assigned to each category (in parentheses). 

\subsubsection{Bot interaction challenges} \label{subsec:interactionchallenges}
Concerning the human-bot interaction on pull requests, the most recurrent and central challenge according to our analysis is that \textbf{bots introduce noise} (12) to the human communication and development workflow. We discuss the results that are specific to noise in Section~\ref{sec:noise}, where we describe the proposed theory about noise caused by bots. 

With regards to bot communication, we unveiled four challenges. We noticed that \textbf{interacting with the bot requires other technical knowledge} (4) not related to the bot. As a consequence, for example, developers might trigger a bot by accident or even misuse the bot capabilities. 
P5 explained that some developers are not aware of how auto-merging pull requests works on GitHub, which leads contributors to misuse the bot that they developed to support this functionality.
This happens due to the way bots are designed to interact. As described by P4, bots perform tasks and need to communicate with humans; however, they do not understand the context of what they are doing. Therefore, we observed that \textbf{bots do not contextualize their actions} (1) and sometimes provide \textbf{non-comprehensive feedback} (3). 
In these cases, when a bot message is not clear enough, developers ``\textit{[...] need to go and ask a human for clarity.}'' [P17], which may generate more work for both contributors and maintainers.
In addition, bots \textbf{do not provide actionable changes} (2) for developers, meaning that some bots' messages and outcomes are so strict that do not guide developers on what they should do next to accomplish their tasks. According to P8 ``\textit{it is great to see `yes' or `no', but if it is not actionable, then it is not useful [...]''}.

Since OSS developers come from diverse cultures and backgrounds, their cultural differences and previous experiences influence how they interact with and react to a bot's action. We observed three main challenges related to developers' \textit{expectation breakdowns} when interacting with bots on pull requests.
First, bots can \textbf{enforce inflexible rules} (4). These rules are commonly imposed by a specific community and evidenced by bot actions. For example, P7 mentioned: ``\textit{so, the biggest complaints we have gotten are that our lint rules and tests are too strict. And of course, the bot enforces that}.'' 
In addition, we found that the way these inflexible rules are interpreted can vary based on developers' expectations. P7 suggested that the bots' ``\textit{social issues largely come down to a bot being inflexible and not meeting somebody's expectations}''.
Another complaint refers to \textbf{bots intimidating newcomers} (3). For new contributors to an OSS project, interacting with a bot that they have never seen or heard of before might be confusing, and the newcomers might feel intimidated, as stated by P12: \textit{``If you're new to a project, then you might not be expecting bots, right? So, if you don't expect it, then that could be confusing''.} Furthermore, some developers might find it \textbf{strange to interact with a bot} (2), as mentioned by P12: ``\textit{`Hey, I'm here to help you' [...] for some people, it is still quite strange, and they are quite surprised by it.}'' Further, P5 also mentioned that receiving ``thanks'' from a non-human feels less sincere.

From an ethical perspective, we identified four challenges. Five participants reported \textbf{bots as intrusive} (5). For them, an intrusive bot is a bot that modifies commits and pull requests: ``\textit{let's say you have a very large line of code and the bot goes there and breaks that line for you. It is intrusive because it is changing what the developers did.}'' [P21]. Another example of intrusive bots are those created for spamming repositories. P4 mentioned the case of \textit{Orthographic Pedant} bot.\footnote{\url{https://github.com/thoppe/orthographic-pedant}} This bot searches for repositories in which there is a typo, then creates a pull request to correct the typo. The biggest complaint about this bot is that the developers did not allow the bot to interact on their projects, as P4 explained: ``\textit{people want to have agency, they want to have a choice. [...] They want to know that they are being corrected because they asked to be corrected.}'' In addition, \textbf{bots impersonating developers} (4) were also mentioned as a challenge by our interviewees. Two other ethical challenges reported during our interviews were \textbf{malicious intent} (2) and \textbf{biased behavior} (2). Bots with a malicious intent could ``\textit{manipulate developers actions}'' [P9], for example, by including a security venerability into the source code by merging a pull request. Further, according to P9, as there is no criteria to verify the use of bots, they can have a \textit{biased behavior} and represent the opinion of a particular entity (e.g., the enterprise who created the bot).

\subsubsection{Bot adoption challenges} \label{subsec:adoptionchallenges}
Participants also mentioned challenges related to the adoption of bots into their GitHub repositories. According to P4, the challenges of bot adoption begin with finding the right bot. Developers complain that it is \textbf{difficult to find an appropriate bot} (3) to solve their problems. As P4 explained, there is a \textbf{limited search mechanism for bots} (3). P6 added: ``\textit{In the [GitHub] marketplace, [...] I don't even know if there is a category for bots.}'' If maintainers find an appropriate bot, they then have to deal with configuration challenges. First, it is \textbf{difficult to tailor configuration} (4) to a project. Even after maintainers spend the time needed to configure the bot, there is no way to predict what the bot will do once installed. In P10's experience, it is ``\textit{easy to install the bot with the basic configuration. However, it is not easy to adjust the configuration to your needs}''. A related challenge is the \textbf{limited configuration} (3) settings provided by the bots. There are limited resources, for example, to integrate the bot into several projects at once. Some participants also mentioned the \textbf{burden to set up configuration files} (2): ``\textit{It is like a whole configuration file you have to write. That is a lot of work, right?}'' [P4]. Maintainers also need to deal with technical complexity issues caused by bot adoption, such as \textbf{handling bots failures} (5). Due to bot instability, our interviewees also mentioned that there is \textbf{more work to monitor bots} (3) to guarantee that everything is working well. Another technical issue is that adopting a \textbf{bot increases the barrier for new maintainers} (1), who need to be aware of how each bot works on the project.

\subsubsection{Bot development challenges} \label{subsec:devchallenges}
We also identified challenges related to bot development. Firstly, bot developers often face \textit{platform limitations}, commonly due to \textbf{restricted bot actions} (2). As mentioned by P5: ``\textit{There are still a few things that just cannot be done with the [GitHub] API. So that's a problem that I face}.'' The platform restrictions might limit both the extent of the bots' actions and the way bots communicate. Regarding the \textbf{restricted bot communication} (2), P4 stated that the platform ideally would provide additional mechanisms to improve it, since the only way bots communicate is through comments. Participants also reported \textbf{technical overhead to host and deploy a bot} (4). P13 identified the ``\textit{main trouble with bots right now is you have to host them.}'' Therefore, when a developer has to maintain the bot itself (e.g., project-specific bots), it becomes an overhead cost, since ``\textit{the bot saves you time but it also costs time to maintain}'' (P19). In addition, we found \textbf{challenges in building complex bots} (4). For example, P12, an experienced bot developer, reports that bots found in other projects ``\textit{just automate a single thing. We just have one bot that does everything. I think it is hard to build a bot that has a lot of capabilities.}''

\MyBox{\textbf{Summary about challenges caused by bots.} We provided a hierarchical categorization of bot challenges and focused on the human-bot interaction challenges. We found 12 challenges regarding bot communication, expectations, and ethical issues. Among these challenges, we found noise as a recurrent and central challenge.}

\subsection{Theory about noise introduced by bots}
\label{sec:noise}

\begin{figure}[!bth]
\scriptsize
\centering
\includegraphics[width=.7\textwidth]{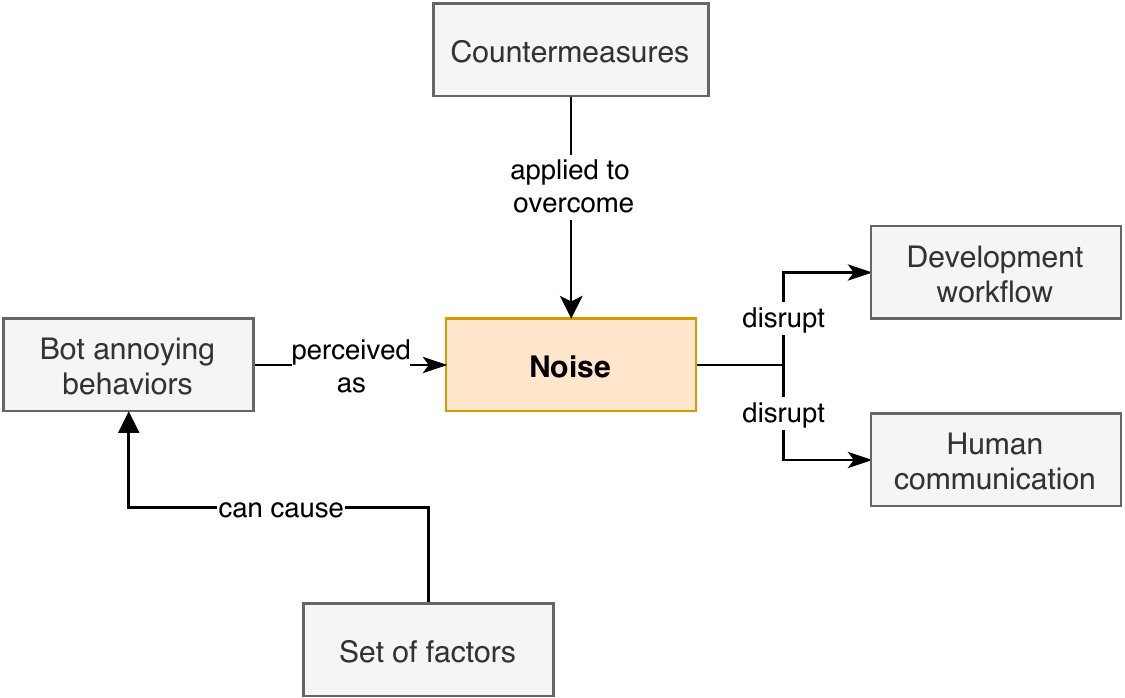}
\caption{High-level concepts and relationships of Bots' Noise Theory}
\label{fig:highlevel}
\end{figure}

As aforementioned, the most recurrent and central problem reported by our interviewees was the introduction of noise into the developers' communication channel. This problem was a crosscutting concern related to bots' development, adoption, and interaction in OSS projects. Figure~\ref{fig:highlevel} shows the high-level concepts and relationships that resulted from our qualitative analysis. Some interviewees complained about \textbf{annoying bot behaviors} such as verbosity, high frequency and timing of actions, and unsolicited actions. Interviewees also mentioned a \textbf{set of factors} that might \textit{cause} annoying behaviors. These behaviors are often \textit{perceived as} \textbf{noise}. The noise introduction leads to information overload (i.e. notification overload, extra information for maintainers), which \textit{disrupts} both \textbf{human communication} and \textbf{development workflow}. To handle the challenges provoked by noise, developers rely on \textbf{countermeasures}, such as re-configuring or re-designing the bot.


\begin{figure*}[!bth]
\scriptsize
\centering
\includegraphics[width=\textwidth]{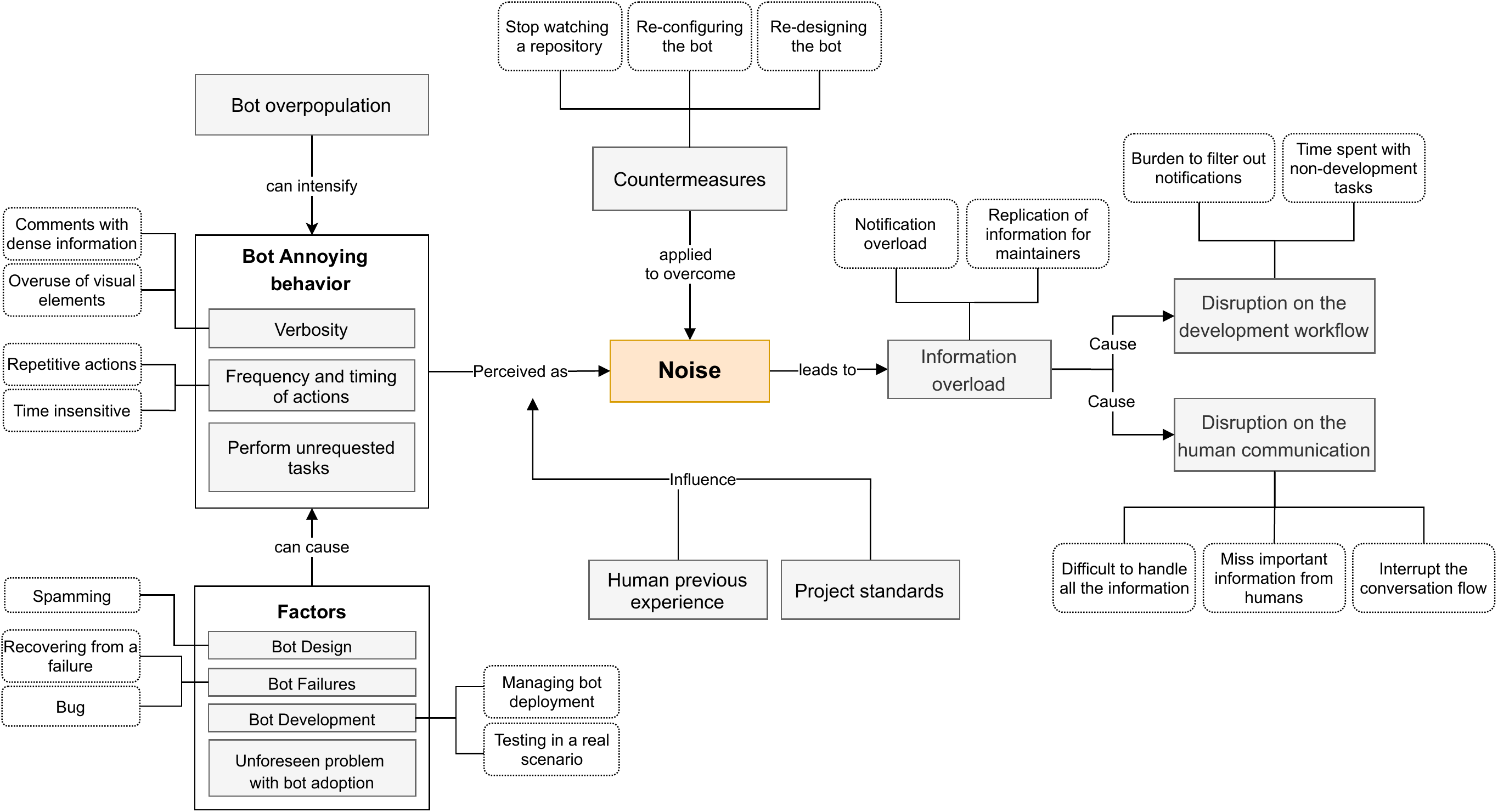}
\caption{Theory about noise introduced by bots}
\label{fig:noise}
\end{figure*}

In the following, we present in detail the theory about noise introduced by bots described in Figure~\ref{fig:noise}. As previously, we include the concepts in \textbf{bold} face and the (sub-)categories in \textit{italic}. We also provide the number of participants for each category (in parentheses). 

\subsubsection{Bot annoying behaviors}

Interviewees reported several annoying behaviors when bots interact on pull requests. The most recurrent one was the high \textbf{frequency and timing} of bots' actions (8). This includes the case in which the interviewees say that bots perform \textit{repetitive actions}, such as creating numerous pull requests and leaving dozen of comments in a row. P6 explained: ``\textit{[...] is an automation that runs too frequently and then it keeps opening up all the pull requests that I do not need or want to.}'' In addition, P9 mentioned complaints about the frequency of bot comments: ``\textit{sometimes we get comments like `hey, bot comments too much to my taste'.}'' Besides that, bot actions are usually \textit{time insensitive}. Bots are designed to ``\textit{work all day long}'' [P10] which might interrupt the developer at the wrong time. P3 offered an exemplary case of how a welcoming bot might be \textit{time insensitive}: ``\textit{as long as, for example, the comment is immediately after I did a change [...] if it is in a second or two and I'm seeing the page I do not get a new notification. But if it happens three minutes later, and I left the page and suddenly I get the new notification and I think `ah, this person has another question or something,' so I need to check it out and find out that this is a bot.}''

Another annoying behavior regards the bots' \textbf{verbosity} (5). Participants complained about bots providing comments with \textit{dense information} ``\textit{in the middle of the pull request}'' [P13], oftentimes \textit{overusing visual elements} such as ``\textit{big graphics}''[P13]. In P19's experience, developers frequently do not like when ``\textit{[...] bots put a bunch of the information that they try to convey in comments instead of [providing] status hooks or a link somewhere}.'' P17 reinforced this issue regarding: ``\textit{[...] a GitHub integration [bot] that posts these rules. Really dense and information rich elements to your pull requests. And I've seen it be a lot more distracting than it is helpful.}'' 

Another common annoying behavior regards the \textbf{execution of unrequested or undesirable tasks} (4) on pull requests. Participants mentioned that, due to external factors, or even due to the way bots have been designed to interact on pull requests, bots often perform tasks that were neither required nor desired by human developers. P6 described an issue caused by an external failure that impacted the bot interaction: ``\textit{Something went wrong with the release process. So, [the bot] opened up a bunch of different pull requests. And like some of them were a mistake. The other engineer that had to comment and be like, `Hey, sorry, these were a mistake'.''}

\begin{figure}[!bth]
     \centering
     \begin{subfigure}[b]{0.32\textwidth}
         \centering
         \includegraphics[width=\textwidth]{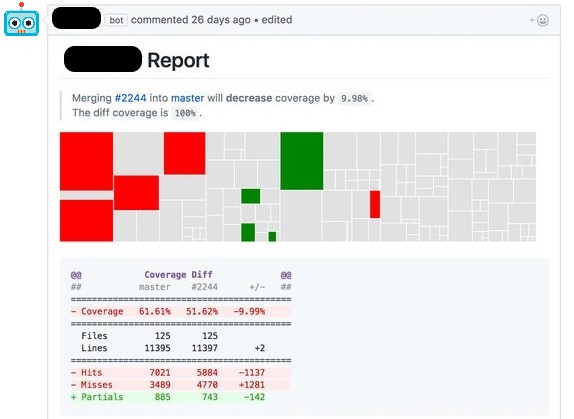}
         \caption{Verbosity}
         \label{fig:verbosity}
     \end{subfigure}
     \hfill
     \begin{subfigure}[b]{0.32\textwidth}
         \centering
         \includegraphics[width=\textwidth]{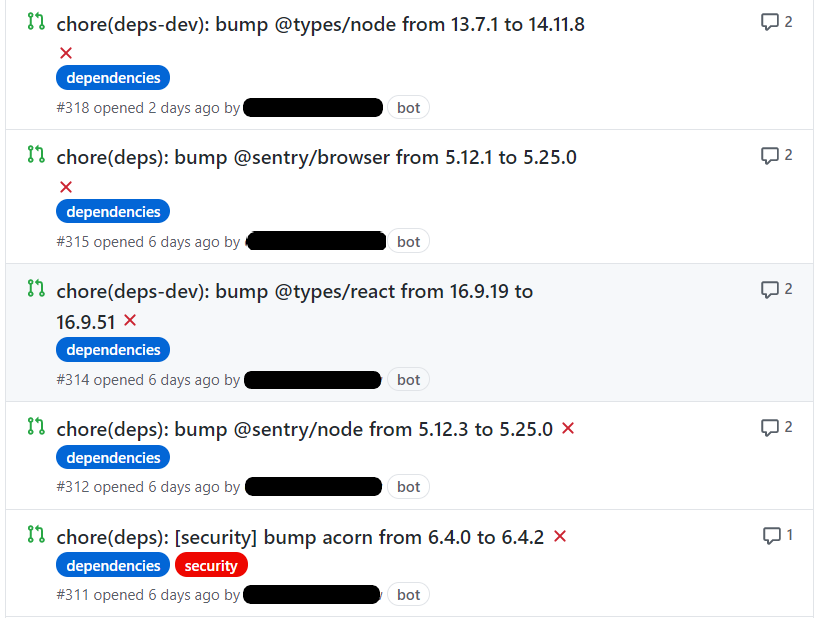}
         \caption{High frequent actions}
         \label{fig:frequency}
     \end{subfigure}
     \hfill
     \begin{subfigure}[b]{0.32\textwidth}
         \centering
         \includegraphics[width=\textwidth]{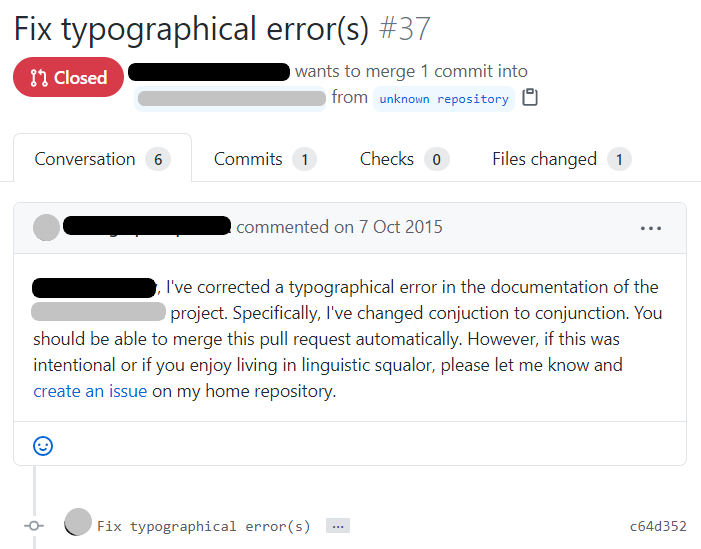}
         \caption{Unsolicited actions}
         \label{fig:spamm}
     \end{subfigure}
        \caption{Examples of annoying behaviors from the state-of-the-practice}
        \label{fig:examples}
\end{figure}

To illustrate the described behaviors, we highlighted some examples cited by our participants and described in the state-of-the-practice. Figure~\ref{fig:verbosity} shows the case of a verbose comment, which included a lot of information and many graphical elements, inserted by a bot in the middle of a human conversation. In Figure~\ref{fig:frequency}, we show a bot overloading a single repository with many pull requests, even if there were opened pull requests by the same bot. Finally, Figure~\ref{fig:spamm} depicts a bot spamming a repository with an unsolicited pull request.

\subsubsection{What might cause an annoying behavior?}

Several \textbf{factors} provoke annoying behaviors, sometimes by \textbf{bots' design} (4). Some bots are intentionally designed to \textit{spam} repositories, as reported in the interaction challenges in Section~\ref{subsec:interactionchallenges}. Other bots might demonstrate a certain behavior by default, as said by P19 when talking about a bot that reports code coverage: ``\textit{but by default, it also leaves a comment}.''

We also found unintended factors that might trigger annoying behaviors. For example, \textbf{bot failures} (4) might be responsible for triggering unsolicited tasks, or even increasing the frequency of bot actions. As shown in Section~\ref{subsec:devchallenges}, handling bot failures is one of the challenges faced by project maintainers. According to P3, when the stale bot, which triages issues and pull requests, recovers from a failure, it posts all missed comments and closes all pull requests that need to be closed. As a consequence, it suddenly overloads both maintainers and contributors. As P7 describes: ``\textit{the only times I perceived our bots as noisy is when there is an obvious bug.}'' Further, an \textbf{unforeseen problem with bot adoption} (3) also may result in unexpected actions or overloading maintainers with new information. Once the stale bot is installed, for example, it comments on every pull request that is open and no longer active. As P3 comments: ``\textit{this is what you want, but it is also a lot of noise for everyone who is watching the repository}.''

In addition, interviewees also mentioned issues during \textbf{bot development} (3) that might trigger an annoying behavior. As reported in the bot development challenges (Section~\ref{subsec:devchallenges}), bot developers, for example, often face technical overhead costs to host and deploy bots. P7 reported that once they tried to upgrade the bot, it led to an ``\textit{edit war},'' resulting in the bot performing unsolicited tasks. Additionally, since there is a lack of test environments for bots under development, bot developers are forced to \textit{test bots in production}. 

\subsubsection{Different perceptions of noise} Bots' verbosity, high frequency of actions, and the execution of unsolicited tasks are generally \textit{perceived as} \textbf{noise} by human developers. This perception, however, might be \textit{influenced by} \textbf{project standards} (3) and \textbf{developers' previous experiences} (3), as noted by P12 ``\textit{what some people might think of as noise is information to other people, right? Like, it depends on the user's role and context within the project.}'' In some cases, for example, an experienced developer may be annoyed by a large amount of information, while a newcomer may benefit from it. As explained by P3, an experienced open source maintainer, ``\textit{when it is your self maintained project, and you see these comments everywhere and you cannot configure the [bot] to turn it off, it might become just noise.}'' For P19, a verbose bot is ``\textit{really more for novices}'' since it ``\textit{tends to have pretty, pretty dense messages.}'' However, dense messages are not necessarily useful for a developer, nor will a new contributor necessarily benefit from them. For P7, newcomers could perceive the bots' verbosity as noise: ``\textit{I do worry that newcomers perceive the bots as noisy, even with only 1 or 2 comments, because the comments are large.}'' In addition, maintainers claim that bots' behaviors might be perceived as noisy when they do not comply with projects' rules and standards. P9 provided an example of this: ``\textit{every public repository has some standards, whether in terms of communication, whether in terms of how many messages the developer should see. And the bot likely will not comply with this policy.}''

\subsubsection{Bot overpopulation} In addition to project standards and developers' previous experience, \textbf{bot overpopulation} (8) might also influence the perception of noise. Eight interviewees reported that annoying bot behaviors can be \textit{intensified} by the presence of several bots on the same repository. As said by P19: ``\textit{because there were 30 different bots, and each one of them was asynchronously going in. So, it was just giving us tons and tons of comments.}'' 

\subsubsection{Effects of information overload}

The bots' annoying behaviors, which are perceived as noise, lead to \textbf{information overload} (7). As stated by P3: ``\textit{it [bot comment] replicates information that we already had.}'' Also, the overload of information can be seen as an \textit{overload of notifications} (e.g., emails or GitHub notifications). It is a problem, as explained by P12: ``\textit{given that we already have a lot of notifications for those of us who use GitHub a lot, then I think that's a real problem.}'' 

Therefore, the information overload negatively impacts both \textbf{human communication} (6) and \textbf{development workflow} (5). Developers mentioned that bots \textit{interrupt the conversation flow} in pull requests, adding other information in the middle of the conversation: ``\textit{\textit{you are talking to the person who submitted the pull request and then a bot comes in and puts other information in the middle of your conversation}}'' [P13]. Participants also mentioned that they usually ``\textit{miss important comments from humans}'' [P1] among the avalanche of information. Due to information overload, it is also hard to \textit{parse all the data} to extract something meaningful. Project maintainers often complain about being interrupted by bot notifications, which disrupts the development workflow. They also started to deal with the \textit{burden of checking} whether it is a human or bot notification. Their time and efforts are also consumed by other tasks not related to development, including reporting spam and excluding undesirable bot comments: ``\textit{I waste five minutes determining that it is a spam}'' [P5].

One practical example of the effects of noise introduction is the case of \textit{mention bot}. The challenges of using this bot were reported in the literature by \citet{Peng2018,Peng2019} and mentioned by P5. Mention bot is a reviewer recommendation bot created by Facebook. The main role of this bot is to suggest to a reviewer a specific pull request. In a project that P5 helps to maintain, a maintainer that no longer works on the project started to receive several notifications when the bot was installed.

\subsubsection{Countermeasures to overcome noise}

We also grouped the strategies that our participants recommend to overcome bots' noise. In most cases, participants reported the \textbf{countermeasures} (6) as a way to manage the noise rather than avoid it. For instance, the noise continues to happen even when a developer \textit{stops watching a repository}. Maintainers also mentioned that they need to \textit{re-configure} the bot to avoid some behaviors. For some bots, it is possible to turn the comments off. During member-checking, P20 reported that in some cases it necessary to \textit{re-configure the bot} to ``\textit{lower the frequency of actions}.'' For example, this is useful for reducing the overload of information generated by dependency management bots, which can submit a couple of pull requests every day. These bots can suddenly monopolize the continuous integration and disrupt the workflow for humans. Another countermeasure that emerged from member-checking is the necessity to \textit{re-design} the bot. P12 mentioned that, after receiving feedback from contributors about noise, they decided to re-design the content of the bot messages and when the bot would be allowed to interact on pull requests.

\MyBox{\textbf{Summary of the noise theory.} We presented a theory of how certain bot behaviors can be perceived as noise on OSS pull requests. This perception often relies on the number of bots on a repository, project standards, and the human's previous experience. In short, we found that the noise introduced by bots leads to information overload, which interferes with how humans communicate, work, and collaborate on social coding platforms.}

\section{Discussion} 

In this section, we discuss our main findings, comparing them with the state-of-the-art. Further, we discuss the implications of this work for the OSS community, bot developers, social coding platforms, and researchers.

Bots on GitHub serve as an interface to integrate humans and services~\cite{Wessel2018,Storey2016}. They are commonly integrated into the pull request workflow to automate tasks and communicate with human developers. The increasing number of bots on GitHub relates to the growing importance of automating activities around pull requests. However, as discussed by \citet{Storey2016} and \citet{Paikari.vanDerHoek:2018}, potentially negative impacts of task automation through bots are overlooked. Therefore, it is critical to understand software bots as socio-technical---rather than technical---applications, which must be designed to consider human interaction, developers' collaboration, and other ethical concerns~\cite{dagstuhl2019}. In this context, our work contributes by introducing and systematizing evidence from the perspective of OSS practitioners who have experience interacting with and developing bots on the GitHub platform.

\paragraph{\textbf{Bot communication challenges}} 
The way bots communicate impacts developers' interpretations and how they handle bot outcomes. According to \citet{lebeuf2018software}, the way bots communicate is important because ``\textit{the bot's purpose -- what it can and can't do -- must be evident and match user expectations.}'' However, we evidenced the necessity of \textit{previous technical knowledge to interact} with and understand the messages of bots on GitHub. Combined with the \textit{lack of context}, it might be extremely difficult for humans to extract meaningful guidance from bots' feedback. These challenges relate to the \textit{platform limitations} bot developers face and the textual communication channel~\cite{botmit}. These findings complement the previous literature, which found that practitioners often complain that bots have poor communication skills and do not provide feedback that supports developers' decisions \cite{Wessel2018}. \citet{Brown2019} argue that designing bots to provide actionable feedback for developers is still an open challenge.

\paragraph{\textbf{Expectations Breakdowns}}
Developers with different profiles and backgrounds have different expectations about bot interaction. Bots, for example, \textit{enforce predefined cultural rules} of a community, causing expectation breakdowns for outsiders. We also found that \textit{bots intimidate newcomers}. New contributors might be confused when interacting with a bot that they have never seen or heard of before. Previous work by \citet{Wessel2018} has already mentioned that support for newcomers is both challenging and desirable. In a subsequent study, \citet{Wessel2020whatexpect} reported that although bots could make it easier for some newcomers to submit a high-quality pull request, bots can also provide them with information that can lead to rework, discussion, and ultimately dropping out from contributing. Developers' different cognitive styles~\cite{gendermag,gendermag2} may also have diverse expectations and their profiles should be considered during the design of bot messages to avoid expectation breakdowns. Differences related to developers' backgrounds are a common cause of problems in distributed software development~\cite{steinmacher2013awareness}. However, when it comes to bots interacting on social coding platforms, it is still an under-explored theme.

\paragraph{\textbf{Ethical challenges}} 
\textit{Intrusive bots} generate ethical concerns. Common intrusive bot behaviors include modifying actions performed by humans, such as changing commits or pull requests content, or even spamming repositories with unsolicited pull requests or comments. Spamming by bots is one of the factors responsible for the perception of noise on GitHub repositories. Another important concern is whether bots are allowed to impersonate humans~\cite{dagstuhl2019}. For bots on Wikipedia, for example, this behavior is expressly prohibited~\cite{muller2013work}. At the same time, \citet{murgia2016among} have shown that individuals on Stack Overflow might be more likely to accept bots impersonating humans as opposed to bots disclosing that they are bots. On GitHub, however, there is no explicit prohibition for \textit{bots impersonating humans}, or even \textit{bots with malicious intent}. Thus, these bots might reinforce stereotypes and toxic behaviors, appear insincere, and target minorities. \citet{golzadeh2020groundtruth} propose a strategy to detect bots on GitHub based on their message patterns. This strategy might be used to identify malicious bots.

\paragraph{\textbf{Noise as a central challenge}}
Noise is a central challenge in bots' interactions on OSS' pull requests. We organized our findings into a theory that provides a broader vision of how certain bot behaviors can be perceived as noise, how this impacts developers, and how they have been attempting to handle it. In communication studies and information theory, the term ``noise'' refers to anything that interferes with the communication process between a speaker and an audience~\cite{mathnoise}. In the context of social coding platforms, we found that the noise introduced by bots around pull requests refers to any interference produced by a bot's behavior that disrupts the communication between project maintainers and contributors.

Although we considered annoying bot behaviors as a source of noise, the perception of such noise varies. Although the overuse of bots potentializes the noise, as also noticed by \citet{erlenhov2020empirical}, we found that noise perception also depends on the experience and preferences of the developer interacting with the bot. For example, while a new contributor may benefit from receiving one or more detailed bot comments with guidance or feedback, an experienced maintainer may feel frustrated and annoyed by seeing and receiving frequent notifications from those verbose comments. Furthermore, noise perception also relies on the differences in the developers' cognitive style and on the limitations humans face to cope with information. For example, Information Processing Theory, proposed by \citet{miller1956magical} in the field of cognitive psychology, describes the limited capacity of humans to store current information in memory. Individuals will invest only a certain level of cognitive effort toward processing a set of incoming information.

A main complaint about noise from developers is the notification overload from bots interrupting the development workflow. Other studies focusing on a single bot also reported that developers can be overwhelmed by bot notifications~\cite{Brown2019,mirhosseini2017can,Peng2018,Peng2019}. According to \citet{erlenhov2020empirical}, there is a trade-off between timely bot notifications and frequent interruptions and information overload. Our findings provide further detail on how developers deal with those notifications and the impacts on the development workflow. Developers deemed notification overload as a significant problem, since they already receive a large number of daily notifications. On GitHub specifically, \citet{goyal2018identifying} found that active developers typically receive dozens of public event notifications each day, and a single active project can produce over 100 notifications per day. The CSCW community for decades has been investigating awareness mechanisms based on notifications \cite{simone1995notation,lopez2016ubiquitous}, which have not been yet explored by social coding platforms. As pointed out by \citet{iqbal2010notifications}, users want to be notified, but they also want to have ways to filter notifications and determine how they will be notified. \citet{steinmacher2013awareness} has performed a systematic literature review on awareness support in distributed software development, which can be used to inspire the design of appropriate awareness mechanisms for social coding platforms.

Our interviewees mentioned the direct effects of information overload on their communication, including difficultly in managing the incoming information and the interruption in the flow of the conversation, which might incur the loss of important information. These effects of information overload have been already observed in teams that collaborate and communicate online~\cite{bawden2009dark,jones2004information,nematzadeh2016information}. According to \citet{nematzadeh2016information}, both the structure and textual contents of human conversation may be affected by a high information load, potentially limiting the overall production of new information in group conversations. In the context of our study, this change in the conversational dynamics described by \citet{nematzadeh2016information} can impact the overall engagement of contributors and maintainers when discussing pull requests. Further, \citet{jones2004information} proposed a theoretical model on the impact on message dynamics of individual strategies to cope with the information overload. According to \citet{jones2004information}, as the information overload grows, users tend to focus on and respond to simpler information, and eventually cease active participation.

\subsection{Implications}

The results of our study can help the software bot community improve the design of bots on ethical and interaction levels. In the following, we discuss how our results lead to practical implications for practitioners as well as insights and suggestions for researchers.

\paragraph{\textbf{Implications for Bot developers:}}

On the path toward making bots more effective for communicating with and helping developers, many design problems need to be solved. Any developer who wants to build a bot for integration into a social coding platform first needs to consider the impact that the bot may have on both technical and social contexts. Based on our results, further bot improvements can be envisioned. One of the biggest complaints about bot interaction is the repetitive actions they perform. In this way, to prevent bots from introducing communication noise, bot developers should know when and to what extent the bot should interrupt a human~\cite{botmit,dagstuhl2019}. In addition, bot developers should provide mechanisms to enable better configurable control over bot actions, rather than just turn off bot comments. Further, these mechanisms need to be explicitly announced during bot adoption (e.g., noiseless configuration, preset levels of information). Another important aspect of bot interaction is the way bots should display information to the developer. Developers often complain about bots providing verbose feedback (in a comment) instead of just status information. Therefore, bot developers also should identify the best way to convey the information (e.g., via status information, comments).

Another point to be considered is that bots spamming repositories was one of the most mentioned ethical challenges by OSS maintainers. It is important for bot developers to design an opt-in bot and provide maintainers control over bot actions. In addition,  our study results underscored that some developers feel uncomfortable interacting with a bot. Human users can hold higher expectation with overly humanized bots (e.g., bots that say ``\textit{thank you}''), which can lead to frustration~\cite{gnewuch2017towards}.

\paragraph{\textbf{Implications for Social Coding Platforms:}} 

Because of the growing use of bots for collaborative development activities~\cite{Erlenhov2019}, a proliferation of bots to automate software development tasks was expected. Recently, GitHub introduced GitHub Actions\footnote{\url{https://github.com/features/actions}}, a feature providing automated workflows. These actions allow the automation of tasks based on various triggers and can be easily shared from one repository to another. However, the way these actions communicate in the GitHub platform is the same as bots~\cite{kinsman2021}, which can lead to the same interaction challenges presented in this study.

Our findings also reveal that there are some limitations imposed by the GitHub platform that restrict the design of bots. In short, the platform restrictions might limit both the extent of bot actions and the way bots communicate. It is essential to provide a more flexible way for bots to interact, incorporating rich user interface elements to better engage users. At the same time, there is a need for well-defined governance roles for bots on GitHub, as already established by Wikipedia~\cite{muller2013work}. Therefore, it is important that bots have a documentation page that clearly describes their propose and what they can do on each repository. Also, it is important to have easy mechanisms so project maintainers can turn off or pause a bot at any time. 

Based on the premise that users would like to have better control over their notifications~\cite{iqbal2010notifications}, GitHub should also provide a mechanism to filter out real notifications from bot ones. This would facilitate the management of bot notifications and avoid wasting developers' time filtering non-humans content. Further, the detection of non-human notifications would help developers identify pull requests that are merely spam.

\paragraph{\textbf{Implications for Researchers:}} 
We identified a set of 25 challenges to developing, adopting, and interacting with bots on social coding platforms. Part of these challenges can be addressed by leveraging machine learning techniques to enrich bots. Thus, we believe that there is an opportunity for future research to support OSS projects by developing smarter bots, thereby providing better human-bot communication. For example, bots could understand the context of their actions and provide actionable changes or suggestions for developers. To design effective bots to support developers on OSS projects, there is room for research on how to combine the knowledge on building bots and modeling interactions from other domains with the techniques and approaches available in software engineering.

Considering that bot output is mostly text-based, how bots present the content can highly impact developers' perceptions~\cite{botmit}. Still, as aforementioned, the developers' cognitive styles might influence how developers interpret the bot comments' content. In this way, future research can investigate how people with different cognitive styles handle bot messages and learn from them. Future research can lead to a set of guidelines on how to design effective messages for different cognitive styles and developer profiles. Further, it is also important to understand how the content of bot messages influences developers' emotions. To do so, researchers can analyze how developers' emotions expressed in comments changed following bot adoption.

Another challenge is related to the information overload caused by bot behavior on pull requests, which has received some attention from the research community~\cite{Wessel2018,Wessel2020,erlenhov2020empirical}, but remains a challenging problem. In fact, there is room for improvement on human-bot collaboration on social coding platforms. Possible future research can leverage noise theory to better support bots' social interaction in the context of OSS. In addition, when they are overloaded with information, teams must adapt and change their communication behavior~\cite{ellwart2015managing}. Therefore, there is also an opportunity to investigate changes in developers' behavior imposed by the effects of information overload.

\section{Limitations and Threats to Validity}

As any empirical research, our research presents some limitations and potential threats to validity. In this section, we discuss them, their potential impact on the results, and how we have mitigated these limitations.

\paragraph{\textbf{Scope of the results:}} Our findings are grounded in the qualitative analysis of data from practitioners who are experienced with bots on the GitHub platform. Hence, our theory of noise introduced by bots may not necessarily represent the context of other social coding platforms, such as GitLab and Bitbucket.

\paragraph{\textbf{Data representativeness:}} 
Although we interviewed a substantial number of developers, we likely did not discover all possible challenges or provide full explanations of the challenges. We are aware that each project has its singularities and that the OSS universe is huge, meaning the bots' usage and the challenges incurred by those bots can differ according to the project or ecosystem. Our strategy to consider different developer profiles aimed to alleviate this threat, identifying recurrent mentions of challenges from multiple perspectives. Our interviewees were also diverse in terms of the number of years of experience with software development and bots. 

\paragraph{\textbf{Information saturation:}} We continued recruiting participants and conducting interviews until we came to an agreement that no new significant information was found. As posed by Strauss and Corbin~\cite{strauss1997grounded}, sampling should be discontinued once the collected data is considered sufficiently dense and data collection no longer generates new information. As previously mentioned, we also made sure to interview different groups with different perspectives on bots before deciding whether saturation had been reached. In particular, we interviewed bot developers and developers who are contributors and/or maintainers of OSS projects. Although we interviewed only 3 contributor-only developers, the analysis of their interviews did not provide new insights when compared to the maintainers who were also contributors.

\paragraph{\textbf{Reliability of results:}}
To increase the construct validity and improve the reliability of our findings, we employed a constant comparison method~\cite{glaser2017discovery}. In this method, each interpretation is constantly compared with existing findings as it emerges from the qualitative analysis. In addition, we also conducted member-checking. During member-checking, participants confirmed our interpretation of the results, requesting only minor changes.

\section{Conclusion}

The literature on bots on social coding platforms report several potential benefits, such as reducing maintainers' effort on repetitive tasks~\cite{Wessel2020whatexpect} and increasing productivity~\cite{erlenhov2020empirical}. In this paper, we investigated the challenges of using bots to support pull requests. We conducted 21 semi-structured interviews with open source developers experienced with bots. We found several challenges regarding the development, adoption, and interaction of bots on pull requests of OSS projects.

Among the existing challenges, the introduction of noise is the most pressing one. Developers frequently complained about annoying bot behaviors on pull requests, which can be perceived as noise. Noise leads to information overload, which disrupts both human communication and development workflow. Towards managing the noise effects, project maintainers often take some countermeasures, including re-designing the bot's interaction, re-configuring the bot, and not watching a repository. Compared to the previous literature, our findings provide a comprehensive understanding of the interaction problems caused by the use of bots in pull requests.

Our study opens the door for researchers and practitioners to further understand the challenges introduced by adopted bots to save developers time and efforts on social coding platforms. For future work, we plan to design and evaluate strategies to mitigate problems related to information overload incurred by the interaction of developers with software bots on social coding platforms, thereby assisting developers to communicate and accomplish their tasks. 

\section*{Acknowledgments}

This work was partially supported by the Coordenação de Aperfeiçoamento de Pessoal de Nível Superior – Brasil (CAPES) – Finance Code 001, CNPq grants 141222/2018-2 and 313067/2020-1, and the National Science Foundation under Grant numbers 1815503, 1900903.

\bibliographystyle{ACM-Reference-Format}
\bibliography{paper}

\end{document}